# IT Security Issues within the Video Game Industry

Stephen Mohr[1] and Syed (Shawon) Rahman, Ph.D.[2]

[1]Information Assurance and Security, Capella University, Minneapolis, MN, USA
`stephenmohr@gmail.com`
[2]Assistant Professor, University of Hawaii-Hilo, HI, USA and
Adjunct Faculty, Capella University, Minneapolis, MN, USA
`SRahman@Hawaii.edu`

### ABSTRACT

*IT security issues are an important aspect for each and every organization within the video game industry. Within the video game industry alone, you might not normally think of security risks being an issue. But as we can and have seen in recent news, no company is immune to security risks no matter how big or how small. While each of these organizations will never be exactly the same as the next, there are common security issues that can and do affect each and every video game company. In order to properly address those security issues, one of the current leading video game companies was selected in order to perform an initial security assessment. This security assessment provided a starting point upon which specific goals and procedures were determined to help mitigate those risks. The information contained within was initially completed on the case study but has been generalized to allow the information to be easily applied to any video game company[1].*

## 1. INTRODUCTION

The video game industry is one of the largest industries around today. As technology advanced, so too have video games as well as security risks. While security risks might not seem like they would be a big issue within the video game industry, recent security breaches prove that security risks can and do plague the video game industry. Numerous video game companies have had security breaches such as Sony, Bethesda Softworks, EVE Online, Minecraft and League of Legends [1] just to name a few.Also the current statistics put the number of American households who play video games is roughly 65%. While this 65% of American households are at risks, the ages of video game players also show that security risks will affect not only the younger generation, but instead can and do affect each and every age group.

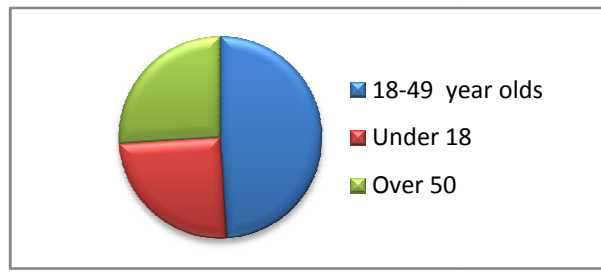

Figure1. Age breakdown of current video game players

---

[1]This work is partially supported by EPSCoR award EPS-0903833 from the National Science Foundation to the University of Hawaii
DOI : 10.5121/ijcsit.2011.3501                    1

In order to properly identify with some real world security issues plaguing the video game industry a leading video game organization was selected to be the case study upon which this article is written.Once the case study was determined a preliminary security assessment was completed. This covers the enterprise as a whole, the potential network vulnerabilities, and even the software and hardware risks that organization would face. The security requirements then cover system security policies, information security training programs, and information security reviews. The final piece to this is a security plan. The security plan details the organizational risks including known risks, priorities, causes, controls, and the potential impacts. An incident response plan is included as well as potential physical security measures and information assurance policies. Also included are the potential implementation plan, the potential barriers, and the potential solutions to those barriers.

## 2. CASE STUDY: ZENIMAX MEDIA INC.

The case study that was selected is ZeniMax Media Inc. They were selected due to them being one of the current leading video game organizations around today. They also currently produce "original interactive entertainment content for consoles, the PC, and handheld/wireless devices" [2]. Most if not all current video game companies produce titles for one or more of those devices which makes ZeniMax Media Inc the perfect case study. Some of their popular titles are the Elder Scrolls series, Fallout 3, DOOM, QUAKE, Wolfenstein, Enemy Territory, RAGE, and Prey.

The current state of ZeniMax Media Inc. is that they are a multinational organization comprised of various subsidies that have been acquired over the years with most of them being within the past two years alone. Those subsidies are Bethesda Softworks, id Software, Bethesda Game Studios, ZeniMax Online Studios, Arkane Studios, Vir2L Studios, Tango Gameworks, and Machine Games. Between ZeniMax Media Inc. and its subsidies, they have offices throughout the US and the world with their main corporate office located in Rockville MD. As for their international offices, they have offices in the UK, Germany, France, Asia, and Benelux which is actually comprised of Belgium, Netherlands, and Luxembourg.

While ZeniMax Media Inc. is the head company now, it was not the actual first. The first company that was created from which ZeniMax Media Inc was spawned was actually Bethesda Softworks which was founded in 1986. It wasn't until 1999 that ZeniMax Media was founded by Robert A. Altman [3]. Since then it has flourished and been a successful video game corporation that can be ranked up there as one of the most prolific and successful video game corporations around.

The purpose of this paper is to first provide the results of a preliminary security risk assessment and then to describe the overall security plan that will address those risks in distinct stages. These stages are the determined security requirements and an extensive security plan that includes specific information assurance policies and procedures as well as the plan on how these will be implemented successfully and effectively to best meet the security needs of ZeniMax Media Inc's enterprise or any other video game organizations that might possibly be in the same situation.

## 3.PRELIMINARY SECURITY ASSESSMENT

A preliminary security assessment is the first step towards developing any information security plan. Risks can come from anywhere within any organization so it is vital to ensure that all areas have been assessed. It begins with the enterprise system's security weaknesses and goes on to discover the system vulnerabilities and exposures on the network and then finally the software and hardware risks. Without gathering the correct information for any and all of these areas could lead to a poor overall security plan. Current security weaknesses and even potential future ones should try to be determined. Each video game corporation will have their own sets of potential security issues that should be addressed and determined.

The following assessment was comprised based upon knowledge obtained through thorough research and is publicly available. Specific information on specific configurations, settings and security risks has been deduced through this research. By performing the security assessment through this manner, it helps to protect the case study from new security risks that could be created through the release of specific business settings and configurations.

## 3.1. Analysis of the enterprise system's security weaknesses

Upon the initial assessment, the biggest enterprise weakness that presented itself is through the process of acquisitions. Acquisitions are a key step in increasing the size of any video game company or any company in general.The amount of acquisitions that are actually involved throughout the video game industry can and does vary but it is still mostly situated around the larger video game corporations similar to the case study.

The current case study has gone through numerous acquisitions already with having four acquisitions in the last year and three the year before.Throughout the actual acquisition process there is a lot of integration involved with the many different systems that could end up creating possible security weaknesses. Anytime you integrate an outside network and system into an existing one, you are bound to run into some level of inherent security risks. Just about every portion of the acquisition process could turn into a security issue. Too much access might be given, domains might not be trusted or valid network traffic might be blocked. Also you could potentially have systems at the new location with known and or unknown security risks. Specific software might have specific risks while others might not be possible to be fully integrated. There are countless possibilities when it comes to acquisitions and the security risks that they might bring.

Most video game organizationsrun a complete MS Windows domain with a fully integrated active directory domain. They also usually would run an Exchange server for their e-mail functions as well. With all of the servers currently running Microsoft Server operating systems, there are risks that can be associated specifically to those types of operating systems. If the systems are not patched regularly there could potentially be security issues. But you can't just go about patching servers blindly because implementing an untested patch could end up creating more risks by breaking production processes, applications, or even potentially crashing the servers. The other issue with using Microsoft servers is that a majority of the known viruses and other malware that is currently floating around on the internet is designed and created specifically to target Microsoft systems.

Throughout the video game industrythere are a variety of programs that could potentially have security risks. The first area where risks are introduced is through the use of commonly used programs that are not only used throughout the video game industry, but throughout just about every company that utilizes computers. Some of the commonly used programs that are exploited are Adobe PDF Reader, QuickTime, Adobe Flash, and Microsoft Office. It is through these commonly used programs that are the primary initial infection vector used to compromise

International Journal of Computer Science & Information Technology (IJCSIT) Vol 3, No 5, Oct 2011

computers that have internet access [4]. Microsoft Office can potentially have security risks if it is not properly configured, but simply having one of the other programs installed could potentially create risks on their own.

For Microsoft Office, certain security measures need to be put into place to ensure that malicious macros and other coding does not run on their own. Microsoft Outlook in particular can have potential security risks as well such as allowing certain types of files to be emailed, etc. Most of these security settings for this specific program would be set on the Exchange server and passed down to all of the clients. As for the other programs, new exploits and vulnerabilities are located fairly regularly and once they are located, they are usually made available to anyone who knows where to look. This is usually done in part to allow people and companies to be aware of potential security risks and what could possibly happen, but it also provides valuable information to hackers by specifically showing them what risks can be exploited and generally how they can be exploited.

One final security weakness that video game corporations will undoubtedly face would be with whatever coding language is being used to create the various products. This is true for any software manufacturer whether it is video game related or regular software related. According to the ISO/IEC TR 24772 [5],

> *All programming languages contain constructs that are incompletely specified, exhibit undefined behavior, are implementation-dependent, or are difficult to use correctly. The use of those constructs may therefore give rise to vulnerabilities, as a result of which, software programs can execute differently than intended by the writer. In some cases, these vulnerabilities can compromise the safety of a system or be exploited by attackers to compromise the security or privacy of a system.*

Basically this states that all programming languages can contain some degree of vulnerabilities, but most of the time the weaknesses and or vulnerabilitiesare not with the language itself but rather with how it is actually written.

## 3.2 Enterprise system vulnerabilities and potential exposures on the network

Video game companies can and do run a wide variety of networking devices that are usually a combination of Cisco and other networking components from various manufacturers. The networks are usually comprised of fully functional TCP/IP LAN networks, while the larger companies would also have a WAN integrated that would connect each of the facilities throughout the world. The current case study has their main headquarters in Rockville Maryland, but they also have other locations within the continental US and even in Europe and Asia[6]. Connecting each of these locations securely can potentially create risks as well if they are not completed properly.

There are a few different ways that video game companies and other corporations can connect WANs securely. One way to interconnect LANs is with a WAN using a dedicated leased line [7]. This is not only a very secure if not the most secure method that could be used to connect WANs, but it is also the most expensive. Because it is the most expensive method, it is often the least implemented method. The other main method would then be to simply connect a company's WANs over the internet. If companies do decide to connect their WANs via the Internet, then the information that traverses it could potentially be compromised if it is not properly secured with proven and effective encryption, authentication, and other security methods.

As with any corporation the case study is comprised of various LANs that have a normal switched network with integrated VLANs. Remote connectivity would most likely be included as well. If an organization does allow remote connectivity then they usually employ a VPN to allow for secure remote connectivity. If not then this would be a security risk that would need addressed. All of these have some degree of security risks. Most of them though are ones that every network faces and it is basically part of the network vulnerabilities that has to be accepted. Some examples of security risks would be if there were no secure safeguards in place such as firewalls, improper configurations, and possibly even the lack of an intrusion detection system or even an intrusion prevention system.

The need for an active intrusion detection system and or intrusion prevention system has become more evident with some of the recent security breaches that have been appearing lately. One prime example is the Sony security breach of 2011. The Sony security breach initially happened between April 16th 2011-April 19th 2011 and exposed a wide variety of information.Within the PlayStation Network (PSN) breach alone, "the lost information included names, addresses (city, state, zip, country), email addresses, gender, birthdates, phone numbers, login names, and hashed passwords" [8]. While an intrusion detection system and or intrusion prevention system would have helped to detect and possibly prevent these intrusions sooner than they were, the breach might not have occurred if other security measures were put into place. According to "Gene Spafford, a professor at Purdue University and executive director of the school's Center for Education and Research in Information Assurance and Security" [9],

> "Individuals who work in security and participate in the Sony network had discovered several months ago while they were examining protocols on the Sony network to examine how the games work, that the network game servers were hosted on Apache web servers-that's a form of software. But they were running on very old versions of Apache software that were unpatched and had no firewall installed, and so these were potentially vulnerable. And they had reported these in an open forum that was monitored by Sony employees, but had seen no response and no change or update to the software. That was two to three months prior to the incident where the break-ins occurred."

This is a prime example on why maintaining proper security protocols is important. By not implementing something as simple as patching a system, can end up costing a company its reputation and even lost revenue. While Sony is a very large organization that has the funds and resources to deal with the consequences of this breach, a smaller organization might not.

Another area that should be addressed is with the acquisitions. With every acquisition there will be a need to open up the network to allow secure access to and from each of the new organizations. Each one needs to have a secure connection to securely allow access,as well as policies and configurations need to be in place to ensure that unauthorized access does not occur. These can be addressed through the use and creation of standard acquisition procedures. Some could be simply to wait until a permanent direct connection is established to each facility or even setting up a semi-permanent VPN connection between each facility. Each time this occurs though there are potential security risks that arise.

One final area that video game companies might face security risk is through potential internet exposures such as websites. Video game companies usually will have various websites for most of their subsidiaries and even for each of their main game titles. Because of this the websites could potentially create security risks as well. First would be if the web servers were not located in a DMZ it could cause issues. Second would be how the web pages are actually made.

Each website could potentially expose the enterprise to security risks. Some examples would be through the use of java script injection, cross site scripting, buffer overflows, and even the error messages that might be displayed could create security risks. If there are large databases that are integrated and do not have proper security measures in place, a hacker might be able to gain access to the information it contains as well. The list goes on for potential security risks that could be created through web pages.

### 3.3 Enterprise systems software risks

Each video game company will have a wide variety of software and hardware in use throughout each company. In regard to the case study, the operating systems and server systems were already discussed within the system security weaknesses along with the security risks associated with utilizing MS Office and Exchange. While these are specific to the case study, they are also some of the most widely used programs throughout all of the video game industry. Each of these systems has inherent risks but with specific policies in place, most of those risks can be mitigated.

While video game programming is similar in nature to software programming, it tends to be quite different. It is different simply because when you start a program it is often static, meaning that all actions require actions from human interaction, while video games can and do run continuously whether there is interaction from humans at all times or not [10]. But when it comes to the core guts of a video game, they still utilize a lot of the same programming languages used to create most software in general.

Some of the programming languages that are used are C++, C#, Java, and even Flash. The next core component would be the game engine. A game engines purpose is to abstract the details of doing common game-related tasks such as rendering, physics, and input [11]. Other aspects of a video game include graphic design, 3D design, scripting, and sounds. Some of the programs are used throughout the video game industry to create the components are Maya, 3Ds Max, and Modo. Scripting is a huge item as well.Without scripts in place, video games would require a lot more specialized coding in place to deal with the repetitive motions that video games are known for.A few of the specific ones are MaxScripts, MEL Scripts, and Python scripts. All of these various specialized programs can and do create risks. Programming language security issues come from poor secure coding practices while the script programs could be used to run powerful automated commands that are extremely important within the video game industry.

Each of these programming languages has security risks that can be integrated into some of the programs that they create, as well as potentially creating risks within the enterprise wherever they are used. One common place for .Net is throughout the various websites that video game company's use. A good example of the potential security issues that websites face recently happened to one of the subsidies of the case study. "According to Bethesda, a hacker group broke into the publisher's websites and grabbed user names, email addresses, and/or passwords" [12]. While no credit card information was stolen, some personal information was stolen right through their website. This was most likely accomplished by using the websites to access either a MS SQL server database and or a SharePoint database. Most large companies today still utilize one or both of these two database types.

Overall there is a wide variety of software that is utilized throughout the entire video game industry. In the case study alone, each of the subsidies could be utilizing a different type of software for creating a video game. You have the 3D software creation programs just like the ones first listed, and then you have the programming languages as well which some have been described so far. Also each console system requires a different programming language. Playstation's have their own distinct programming language as well as Xbox has a distinct one as well.

### 3.4 Enterprise systems hardware risks

As for the hardware that is in use there could be a wide variety of systems throughout each video game company. The servers that are in use though can come from Dell, HP, or even IBM to name a few. The current case study utilizes Dell servers with a couple of HP servers within their main corporation. The reason that servers are usually purchased through these computer suppliers is that they usually include a warranty that allows for parts to be easily replaced free of charge. This in turn reduces a company's potential hardware risk should a component fail. UPS's are also employed to protect the servers and workstations, but if one fails then the servers could then be at risk from spikes in the power and brownouts which could damage hardware in use.

Throughout the video game industry, there are also a wide variety of workstations that could be utilized. Depending on the type of work that an employee performs can be the determining factor upon which system to utilize. The graphic designers and 3d programmers might utilize an Apple system, while the programmers and the administration staff might use a normal PC. The current case study employs Dell workstations and even Dell laptops throughout most of their enterprise. A majority of these workstations are kept the same model to help ensure that a standard secure image can be placed on them all. If there was a wider variety of workstation with only one or two models here and there then it would create risks for the organization. Having a standard workstation or multiple workstations throughout a company reduces the potential for a security risk to be introduced. Standard images help to ensure that all systems will have the correct patches, updates, and settings on each workstation.

## 4. SECURITY REQUIREMENTS

Based upon the security assessment the current system security needs are a good solid information security training program and an information security review program as well as the creation and implementation of various standards and policies. These will provide a solid first line of defenceand can provide the appropriate security measures to help ensure that the training programs are both effective and efficient. All of these are requirements that should be included throughout each video game company within the video game industry as they have been proven time and again to provide the best system security needs.

### 4.1 System Security Policies

The first part to be introduced would be the creation of specific system security policies. There is a wide variety of possible policies that could be created and implemented, but the main ones that would need to be created would be an enterprise information security policy (EISP), an issue-specific security policy (ISSP), and a systems-specific policy (SysPS).

An enterprise information security policy is the first line of system security policies that any organization would need. The purpose of it is to provide "an overview of the corporate philosophy on security" [13].On top of that it is also a way to lay out the details on what all responsibilities are required to ensure the security of each organization. This is especially important for larger enterprises that need the specific responsibilities determined beforehand.

An issue-specific security policy specifically covers topics such as authorized access, equipment usage, and even systems management. Each of these is specific issues that can and should be addressed. They are all important specific policies every video game company needs to address. Other areas that would be included could include an e-mail policy, a use of internet policy and even a specific minimum configuration policy. One final policy that should be present would be a social networking policy. This is especially important in today's world where social networks are used throughout the entire video game industry.

The systems specific policies can cover a wide variety of areas. Some are managerial guidance specifications, while others can be technical specifications. Basically anything that would need to be specifically addressed throughout each company would be included within one of these SysPS's. All three of these policies combined should cover any video game company in regards to overall security policies. In any case each of these will need to be specifically created around each video game company and their systems as well as throughout all of their subsidies if there are any present.

### 4.2 Information Security Training Programs

The next area that needs to be included would be the information security training programs. Training programs need to be created around three distinct elements. Those distinct elements are security education, security training, and security awareness. Each of these wouldcover a different area within each company.

Security education will provide insight and understanding into the specific security policies that are designed around each organization and their subsidies. Security training will provide the knowledge and skills necessary to provide employees with the tools and skills necessary to prevent and address security risks. Security awareness then provides valuable information on potential exposures that employees might face. This could be both the normal e-mail security risks, and even the more specific security risks such as the software and programming risks that the programmers would mostly face.

### 4.3 Information Security Reviews

Information security reviews are the final area of information security requirements. These are necessary to ensure that any of the policies designed and implemented that were discussed earlier, are actually effective and working. Without reviewing the security controls and policies, there would be no way to ensure that they are being implemented effectively until a security incident occurs. By that time it would be too late.

Information security reviews can be completed in a few ways. One of the more common types would be simply having logs created that monitor the network and other important areas for access attempts and even failures. Another area would be giving employees a test or utilize some other type of methods such as requiring a report can help show that the training that is being provided is actually being learned by the employees. There are many different ways overall that this can be completed and depending on what all policies and training programs are actually introduced, can be a determining factor in how they are reviewed.

### 5.0 THE ROLE OF TRAINING

The role of training throughout the video game industryis very important.First of all the more common types would be ensuring that employees know how to handle their passwords securely, and secondly, even how to act securely while accessing the internet from within the corporation. These two are basic ones that every organization should have but they are extremely important. Without formal training, it would be hard to ensure that users know how to handle themselves securely while working.

The next area would be to train the employees on the reasons why it is important to act securely and how to react securely in the face of a security incident. This should also cover specifically how to cover the acquisitions securely. These both would provide insight into why information security is important, and it can also provide some of the knowledge necessary to react securely if necessary. Another type of training that should be introduced would be how to spot security risks. The employees are the first line of defence and the only way that this is possible is by ensuring that users know what to look for.

Other training that should be provided to the programmers is ways to ensure that the software they produce is secure and as free from flaws as possible such as through secure coding practices and even improved coding methods. This is especially important because when working with the code of programs, one could easily create a gateway into the network, whether intentionally or accidentally. By training the programmers on how to program securely, video game companies can ensure that the products they are producing are safe and secure in both the development process and final distribution.

### 5.1 Training Materials

In order to properly train the various employees throughout the video game industry, there are a few different types of training materials that can and should be employed. First off if a company and its subsidies are spread throughout the world such as the case study, there needs to be a way that training can be completed by all of the employees no matter where they are. One such method that was determined to be a good training material would be through the use of an independent study or online training session that could easily be accessed by any employee internally. This might not be used for all training programs, but it can be efficient for training employees on how to handle themselves securely online, or how to handle their passwords securely, or even how to securely handle e-mail messages. All of these general policies should be taught to all employees. Also newsletters can be a good method or even videos would work.

As for the more specific training programs that would be introduced such as the programming securely training could take place via lecture. This helps to provide more hands on interaction which usually is more easily remembered. Plus with the lecture type of setting, employees can ask specific questions or even provide student to student expertise that others employees have found to be effective. If lectures can't be implemented at all facilities and locations, then the lecture could in turn be presented within an online course room setting that allows for one-on-one interaction between employees and the lecturer.

## 6.0 SECURITY PLAN FOR VIDEO GAME INDUSTRY

### 6.1 Organizational Risks

Organizational risks range from the integration of the networks and systems of the many acquisitions that are completed, to the potential software and hardware risks that come from the both the implementation of new and current systems, as well as through the creation of the new video games. Each of these brings a certain level of risks to the enterprise as a whole, which is the exact reason for the necessity of prioritizing those risks so that they are addressed in a specific order.

The following table includes all of the various risks that were determined within the preliminary security assessment report that was completed on the case study. Within this table are specific information on each risk such as the priority of the risk, what the potential risk is, what the cause of the risk is, as well as the potential controls for each risk and its potential impacts on the enterprise as a whole.

Table 1.Risk, threats, priorities, controls, and impacts.

| Priority | Risk or threat | Cause | Potential controls | Impact |
|---|---|---|---|---|
| High | Integration risk, domain integration risk, and software integration risks | Acquisitions, mismatched network\software systems, poor\improper configurations, relaxed security rulesensure integrations run smoothly and successfully | Predefined acquisition procedures that includecompleting a full inventory of the acquisitions systems and settings. Research can then be complete to determine a precise integration plan. | The potential impacts can be tremendous due to the amount of information and systems that would need integrated into existing systems |
| High | Server systems crashesand unexpected but avoidable issues | Untested upgrades and patches hastily installed can cause system crashes, new security risks, etc. | All upgrades\patches should be thoroughly tested within a development system that to properly gauge if there are any potential issues before being introduced into the production environment. | Server & workstation downtimes, or other unforeseen consequences |
| High | Portable devices expose corporate information or allow for an entry point into the network | Improper security configuration, non-existent portable security policies, use of public unsecured internet access, improper storage of hardware when not in use, etc. | Portable security policy creation, secure standardized image configurations for any and all portable devices,prevention of unauthorized portable devices, training on remote best use practices. | Large or small depending mainly on the type information is stored on a device and what type of access it might provide to the end user. |
| Medium | Risks associated from websites | Cyber attacks from external or internal threats through the use of varying | Websites should be placed into a DMZ to ensure that there is an extra layer of security between the websites and anything | Company websites are for informative reasons only but can still provide an entry point into |

International Journal of Computer Science & Information Technology (IJCSIT) Vol 3, No 5, Oct 2011

| Priority | Risk or threat | Cause | Potential controls | Impact |
|---|---|---|---|---|
| | | attacks such as DOS\DDOS attacks, script injection, cross site scripting, etc. | internal. Also a possible intrusion prevention system should be implemented to help prevent some of the potential attacks that might occur. | the network if not properly configured |
| Med-Low | Interception or alteration of information from connection between various locations | Unless there is a direct connection between facilities, then information that is transferred between connections could potentially be hijacked or intercepted. | Direct connections to each of the facilities should be installed whenever possible. If that is too costly or unable to be efficiently completed, then proper encryption protocols should be enforced to properly encrypt the transmissions. | If proprietary information such as the source code to one of their products is intercepted then it could lead to lost revenue, etc. |
| Low | Programming risks from specific programming languages | This can be caused by not using secure coding practices and by using outdated programming languages, methods, etc. | Utilize current industry best secure coding practices and thorough scans also should be completed to scan for common programming flaws that might otherwise be missed. | The potential impact for these will vary depending on each type of vulnerability or risk. |

### 6.2 Incident Response Plan

If any of these incidents do occur, then they will need to be responded to in an orderly defined fashion. This is where an incident response plan is necessary. The following are generalized information or steps on how to properly respond to an incident. Since they are generalized, they can be molded into specifically to each specific situation.

### 6.2.1 – Generalized information response plan

1. "Users of information services are required to note and report any observed or suspected security weaknesses in, or threats to, systems or services immediately upon observation" [14].
2. Users should wait until after an IT/security specialist verifies whether or not an incident has actually occurred.
3. A designated IT/Security specialist will then be responsible for checking out an incident and documenting its cause, severity, and its current impact on the company.

4. Once the initial information is documented, then each specific incident can be addressed and mitigated.
5. Then finally once the incident has been mitigated, it needs to be documented by the IT/Security specialists on what controls were implemented.
6. A post analysis should be complete on the documentation that was acquired to learn from it in order to better address each incident or what steps can be taken to mitigate the risk entirely.

### 6.3 Physical security measures

The next portion of this security plan will be to address any of the physical security measures that should be utilized. The physical security measures that should be installed if they are currently missing would start at the server room locations. Each server room should be equipped with working environmental controls that keep the server rooms cooled to an appropriate level. Also any location where there are severs located should be equipped with the appropriate fire systems to best protect the servers and equipment. The best type of automatic fire suppression system for electronic and server room equipment would be a CO2 system. This is a great system to use around electronics because "carbon dioxide extinguishes a fire by removing its supply of oxygen" [15], versus other methods that could potentially destroy electronics. However because it removes all of the oxygen from a room it could end up suffocating a person should they be in the room when it goes off. Because of that there can be installed personal oxygen tanks present or some other way of protecting a person should they be stuck in the server room when it goes off. The rest of the facilities would need to have the normal water type systems that would not harm the employees.

The next portion would be installing an alarm system that is equipped with motion controls and live feed cameras in all of the server rooms and other important locations. These live feed cameras should be recorded and monitored at all times by an appropriate security guard. Other physical security measures would then include the use of a smart UPS that can protect the servers from brownouts, spikes, or simply from being shut off unexpectedly which is known to cause corruption and other damage to the data stored on them.

The final area within the physical security area would be to have all of the employees carry a badge with a smart chip in them that are programmable and will allow users into areas that they have been approved to be in. These badges, or key cards, will be utilized in conjunction with a unique password that users will be required to change every 90 days. Other reasons for having the badges is it provides a second layer of security on top of the normal password that most people and companies are accustomed to. They will use them to log into their computers and to unlock specific doors into secured locations. The users will also be required to have their badges with them at all times while on the facility. Without it, then they will not be able to log into the network or even get into the specific areas that they might need to work.

### 6.4 Information Assurance Policies

In order to provide effective information assurance policies to any video game company, there are a few areas that need to be determined. One area is through the use of redundancy. The first step to making these systems redundant are through the use of a RAID 5 system in each of the servers. This provides the initial redundancy because if one of the drives crash then a new one can be used to replace the defective one. This is very useful and important within every video game company because it also allows for the systems to remain operational as long as possible. If one drive goes down in a RAID 5 the system will still remain functional.

The next area is through the use of cloned servers at multiple locations. The most common type of cloned server would be the domain controllers. Each of the locations, or at least the main locations, should have a redundant domain controller. This is beneficial because it allows users to log in not matter which domain controller is accessed in order to authenticate them. File systems might also be cloned as well. Having the working material located in two different areas provides redundancy as well in not only by backing up the data, but also by making it more available or accessible in the event that the network connection to the main location happens to be broken. All of the time there are circumstances that could occur that might cut off one of the facilities from the rest of them.

The final portion to this plan is to enable a reliable backup solution. The backup solution that is recommended is to first classify all of the data appropriately. Once the data is classified appropriately then the backups can be determined. Basically though there should be full backups completed each week. Throughout the week there will be a combination of both incremental and differential backups. The differential backups could be completed on the important information that would need to be restored the quickest, while the incremental backups would focus more on the second hand data that can be restored after the main data has been.

To store all of these backups all of the weekly incremental and differential backups and at least three full backups should be stored on-site within environmentally controlled and fire protected safes. At the end of each month, one of the full backups should then be transferred off-site into long term storage vaults that are also environmentally controlled and fire proof. This will help ensure that no matter if one facility has a disaster and the backups on site were not recoverable, there will always be a backup that can be retrieved from the long term vault.

All of these backups will need to be completed by the appropriate roles at the main server locations. The specific roles that will be involved will be responsible for ensuring that the backups run as intended and that when new data is presented that needs backed up, then those roles will be able to include the new information into the weekly rotation. They will also be responsible for storing the backup tapes on-site and then off-site as well.

### 6.5 Initial implementation plan

To implement these plans successfully throughout any video game company,you will first need to have all of the potential roles who will be involved throughout each implementation determined beforehand. Once the specific roles are determined, then each employee who fits those roles will need to be educated on how to properly implement these plans. The most important role will be the CIO who will be responsible for working with both the technical side and business side of each new procedure and policy to ensure that they will fit and meet the needs of each video game company. The other roles would be then the manager, and IT staff who will need to work with all of the employees to coordinate and implement the plans effectively and efficiently.

Once the roles and responsibilities are determined, a thorough inventory of each location will need to be completed that includes both the physical conditions and physical security measures that are already in place. By knowing what all is present throughout each company and the locations of each, any new security plans and policies that will be created can be crafted and molded to best suite each company independently.

## 6.6 Potential Barriers

The potential barriers that might come from implementing this security plan come mainly from the acquisitions. Each acquisition will have a different network already in place. Because of that, not all networks can function securely together without first performing a major overhaul of their current setups. Also employees will most likely be tough to get to properly take on any new procedures and policies if they go against what they have been doing previously. Employees can sometimes be the biggest barrier simply because most people do not like change. They will take the easiest approach if possible in order to prevent any changes.

## 6.7 Potential barrier solutions

Some solutions to these potential barriers are to mainly educate and train employees on any new changes. As the new changes take effect, another method would be to provide incentives for following any new security plan. The last portion of the barrier is to also research as much as possible in order to best integration and merge any new company. The complete inventory will allow the proper roles to research and see what will work together and what might need to be removed. These were only just a few of the potential solutions, but they do help mitigate most if not all of the potential barriers that might be discovered.

## 7. CONCLUSION

The current video game industry is extensive in size and is continuing to grow. New companies are created, sold, acquired, and even closed. IT security concerns are a crucial aspect for each and every organization within the video game industry. Through the successful creation and implementation of any enterprise security plan, each and every one of those companies will be able to maintain the proper security and infrastructure to maintain a competitive place within today's video game market. The information provided should provide most video game companies a solid starting point in creating and implementing a security plan more specifically centered on each company. This as well as any and every security plan that is designed and created is a work in progress. The information is current as of the date of this article, but a year or more from now there may be components that should be replaced or upgraded. New security procedures arise, new hardware is created and purchased, new software is produced, and even new business ventures are acquired. The world is always advancing and changing and the video game industry is no different. Addressing the security issues and implementing the security controls today, will only strengthen each company tomorrow.

International Journal of Computer Science & Information Technology (IJCSIT) Vol 3, No 5, Oct 2011

**Authors Bio:**

**Stephen Mohr** is an IT Professional currently pursuing a career within the information security field.  Stephen Mohr's current academics include an Associate's degree specializing in Computer Forensics, as well as a Bachelors Degree specializing in Information Assurance and Security.  His current interests include computer forensics, information assurance and security, networking, computer system designs, current security trends, video games and new technology in general.

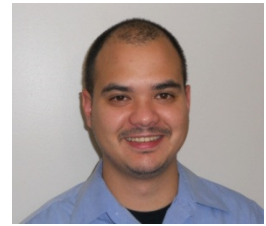

**Dr. Syed (Shawon) M. Rahman** is an Assistant Professor in the Department of Computer Science and Engineering at the University of Hawaii-Hilo and an adjunct faculty of information Technology, information assurance and security at the Capella University. Dr. Rahman's research interests include software engineering education, data visualization, information assurance and security, web accessibility, and software testing and quality assurance. He has published more than 70 peer-reviewed papers. He is a member of many professional organizations including ACM, ASEE, ASQ, IEEE, and UPE.

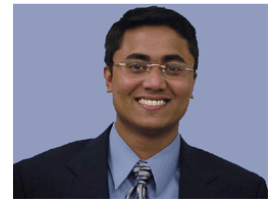